\documentstyle[preprint,eqsecnum,aps]{revtex}
\font\eu=eufm10
\begin{document}
\draft

\title{Jordan-Wigner Transformations and Their Generalizations for
Multidimensional Systems}

\author{Martin S. Kochma\'nski\\}
\address{Institute of Physics, 
University of Rzesz\'ow\\
T.Rejtana 16 A, 35--310 Rzesz\'ow, Poland\\
e-mail: mkochma@atena.univ.rzeszow.pl}
\date{\today}
\maketitle

\begin{abstract} 
In the paper nonlinear transformations of the Jordan-Wigner (JW) type are
introduced in the form different from the ones known previously, for the 
purpose of expressing multi-index Pauli operators in terms of multi-index 
Fermi creation and annihilation operators. These JW transformations in the 
general case being a subject of a rather complicated algebra of transposition 
relations  between various sets of Fermi creation and annihilation operators, 
depending on the common multiindex  of the latter, is shown. As an example, 
the two- and three- dimensional transformations of the JW type are 
investigated, their properties and possible applications in analysis of a 
couple of lattice models of statistical mechanics and also an example of 
application of these transformations to problems of self-avoiding walks in 
graph theory, are discussed. The relation of the obtained transformations to 
the previously known transformations  of the JW type for higher dimensions is 
shown.
\end{abstract}

\section{Introduction}

In 1928 P. Jordan and E. Wigner have introduced their celebrated 
transformations \cite{1}, which enabled expressing Fermi operators $c^{\dag}_m$ 
and $c_m$ in terms of the Pauli operators for a one-dimensional system.  
Namely, the following relations are satisfied (we use the notation generally 
accepted in contemporary literature on the subject):
\begin{equation}
c_m=\exp{\left( i\pi\sum^{m-1}_{j=1}\tau^+_j\tau^-_j\right)}\tau^-_m,\;\;\;
c_m^{\dag}=\exp{\left( i\pi\sum^{m-1}_{j=1}\tau^+_j\tau^-_j\right)}\tau^+_m,
\end{equation}
where $\tau^\pm_m$ - Pauli operators \cite{12}, satisfying anticommutation 
transposition relations for one node $m$ of the chain
\begin{equation}
\left\{\tau^+_m,\tau^-_m\right\}_+=1, \;\;\;\;\;
\left(\tau^+_m\right)^2=\left(\tau^-_m\right)^2=0 ,
\end{equation}
and commutation transposition relations for different nodes of the chain
\begin{equation}
\left[\tau^\pm_m,\tau^\pm_{m'}\right]_-=0\hspace{1cm} (m\neq m').
\end{equation}
The Pauli operators $\tau^\pm_m$ can be expressed in terms  of the Pauli 
matrices $\tau^k_m$, $(k=x,y,z)$ \cite{18,Baxter}  by the following 
well known formulae:
\begin{eqnarray*}
\tau^{\pm}_m =\frac{1}{2}\left(\tau^z_m\pm i\tau^y_m\right), \;\;\; 
\tau^x_m =-2\left(\tau^+_m\tau^-_m-\frac{1}{2} \right), \;\;\;
\tau^z_m =\tau^+_m+\tau^-_m .
\end{eqnarray*}
There exist also transformations that are inverse to (1.1):
\begin{equation}
\tau^-_m=\exp{\left( i\pi\sum^{m-1}_{j=1}c^{\dag}_jc_j\right)}c_m, \;\;\;
\tau_m^+=\exp{\left( i\pi\sum^{m-1}_{j=1}c^{\dag}_jc_j\right)}c^{\dag}_m
\end{equation}
and the following relation is satisfed: $\tau^+_m\tau^-_{m}=c^{\dag}_mc_m$.

The Jordan-Wigner (J-W) transformations (1.1), (1.4) are applied widely in 
numerous domains of  quantum as well as  classical physics, especially in 
quantum field theory, statistical mechanics of quantum and classical 
systems, in physics of phase transitions and critical effects 
\cite{12,2,4,5,6,7,9,10,11,Foda}, etc. One of the  most spectacular 
examples of application of Jordan-Wigner transformations is the outstanding 
paper by Schultz, Mattis and Lieb \cite{11}, in which the authors introduced  
a new approach to solving the two-dimensional Ising model. The key moment of 
the paper is application of the J-W transformations (the transfer 
matrix method was known earlier \cite{2}), and then reduction of the problem 
to a problem  of many fermions  on a one-dimensional lattice, i.e. 
transformation to the fermion representation. The solution \cite{11} of the 
Lenz-Ising-Onsager problem is, in our opinion, the  most beautiful and 
powerful of  all solutions  given before and after this one. It seems to be 
the proper place to mention the note made by J. Ziman \cite{14} p. 209, which 
sounds a little bit pessimistic. The note  concerns the question  connected  
with various approaches to the solution of the Lenz-Ising-Onsager problem 
after switching on external magnetic  field of a finite value. Ziman, 
reporting  the approach introduced  in the paper \cite{11}, writes that 
taking into account external magnetic field extremely complicates the 
operator representation of the transfer matrix in the second quantization 
representation and that {\it "Exactly in this point the limitation of the 
Onsager's method is manifested"}.

We feel that there is some misunderstanding  in this statement, because using 
the Onsager's method we do not apply the field-theoretical language  of 
creation and annihilation operators for fermions (or bosons), but the authors 
of the paper \cite{11} do. Indeed, the Onsager's method \cite{15}, \cite{16} 
exhibits some limitation concerning its application to solving the 
two-dimensional Ising model in external magnetic field or in solving the 
three-dimensional Ising model \cite{18}. In contrast, we deal with a 
completely different situation considering the approach of Schultz, Mattis 
and Lieb \cite{11}, which uses with all of its beauty the field theoretical 
language of second quantization. This method should no longer be treated as a 
beautiful trick but as a powerful tool with great prospects for 
generalizations. The first step towards such a generalization,  leading to the 
solutions of the Lenz-Ising-Onsager problems for $d=2$, $H\neq 0$ and for 
$d=3$, $H\neq 0$, is (in additionally to the transfer matrix method) the 
introduction of the J-W transformations, generalized to two-dimensional and 
three-dimensional systems (see \cite{25,26}). 

Further in this paper the transformations (1.1) and (1.4) will be called 
one-dimensional Jordan-Wigner  transformations. It is well known that
instead of  transformations (1.1), (1.4) we can introduce  transformations of 
the form:
\begin{equation}
\label{6a} b_m=exp\left(i\pi\sum^{\mbox{\eu m}}_{p=m+1}\tau^+_p\tau^-_p
\right)\tau_m^-\hspace{1cm}b_m^{\dag}=exp\left(i\pi\sum^{\mbox{\eu m}}_{p=m+1}
\tau^+_p\tau^-_p\right)\tau_m^+,
\end{equation}
and the transformations inverse to them. These transformations will be called
inversional. Operators $(b^{\dag}_m,b_m)$ are obviously the Fermi creation 
and annihilation operators. Operators $(c^{\dag}_m, c_m)$ and  $(b^{\dag}_m,
b_m)$  are connected by relations  of the type:
\begin{equation}
\label{6b}  c^{\dag}_m=-(-1)^{\hat{\mbox{\eu m}}}b^{\dag}_m,\hspace{.5cm}
c_m=(-1)^{\hat{\mbox{\eu m}}}b_m,\hspace{0.5cm} c^{\dag}_mc_m=b^{\dag}_mb_m,
\hspace{0.5cm}m=1,2,\ldots,\mbox{\eu m}
\end{equation}
where: $\hat{\mbox{\eu m}}=\sum_m c^{\dag}_mc_m=\sum_mb^{\dag}_mb_m$ - the 
operator of the number  of all fermions, as it follows from (1.1), (1.4)-
(\ref{6a}). It is easy to show that 
\begin{equation}
\label{6c} [c^{\dag}_m,b^{\dag}_{m'}]_-=\ldots=[c_m,b_{m'}]=0,
\hspace*{.8cm} m\neq m',\hspace{1.5cm}[c^{\dag}_m,b_m]_-=-(-1)^{\hat{\mbox
{\eu m}}}, 
\end{equation}
i.e. these operators  commute for different $m$.

Recently appeared a few papers \cite{20,21,22,23}, investigating 
generalization of J-W  transformations   for lattice systems to higher 
dimensions. Fradkin \cite{20} and Y.R. Wang \cite{21} consider  
generalization to the two-dimensional case (2D), and Huerta and Zanelli 
\cite{22} and S. Wang \cite{23} - to the three-dimensional case (3D). The 
latter two authors show also that appropriate generalizations to the 4D and 
higher dimensional cases is straightforward. Further in the paper we refer to 
some results of the papers \cite{22,23} and especially from the paper of S. 
Wang \cite{23}. Therefore let us remind shortly the most important points of 
the paper. In the paper \cite{23} were given solutions to the equations 
(here we preserve original notation of the author):
\begin{equation}
S^-({\bf x})=U({\bf x})c({\bf  x}), \;\;\;
S^+({\bf x})=c^{\dag}({\bf x})U^+({\bf x}),\;\;\;
U({\bf x})=\exp{\left[i\pi\sum_{\bf z}w({\bf x,z})c^{\dag}({\bf z})c({\bf
z})\right]}
\end{equation}
where $S^{\pm}({\bf x})$ - Pauli operators, $c^{\dag}({\bf x}), c({\bf x})$ - 
Fermi operators, and the function $w({\bf x, z})$ should satisfy the 
following condition:
\begin{equation}
e^{i\pi w({\bf z,x})}=- e^{i\pi w({\bf x, z})},\hspace{1cm}{\bf x}\neq{\bf z}
\end{equation}
The solution of the equation (1.9) is of the form:
for the 1D case $w(x,z)=\Theta (x-z)$, where $\Theta (x)$ -- the step-like
function (the unit - valued Heaviside function), and for the 2D case
\begin{equation}
w({\bf x, z})=\Theta (x_1-z_1)(1-\delta_{x_1z_1})+\Theta (x_2-z_2) \delta_
{x_1z_1}
\end{equation}
where $x_{1,2}$ and $z_{1,2}$ - components of the vectors ${\bf x}$ and 
${\bf z}$, respectively, in a chosen coordinate system $({\bf e}_1,{\bf e}_2)$, 
and $\delta_{xz}$ -the one-dimensional lattice delta function. Finally, for 
the 3D case S. Wang \cite{23} writes the solution:
\begin{eqnarray}
w({\bf x , z})&=&\Theta (x_1-z_1)(1-\delta_{x_1,z_1})+\Theta
(x_2-z_2)\delta_{x_1,z_1}(1-\delta_{x_2z_2})\nonumber\\ &+&\Theta
(x_3-z_3)\delta_{x_1z_1}\delta_{x_2z_2}
\end{eqnarray}
It is easy to see that the role of the step-like  function $\Theta (x)$ could 
be played by one of the three Heaviside functions: $\Theta_s (x)$ - the 
symmetric unit-valued function, $\Theta^{\pm}(x)$ - the asymmetric unit-valued 
functions. One should  deal with $\Theta_s (x)$ with a special care.

We will not explore here various topological aspects of the $w({\bf x,z})$,  
that were briefly discussed in the papers \cite{20,21,22} for the 2D case, 
and in the paper \cite{23}  for the 3D case, although from a different point 
of view. More specifically, in the paper \cite{22} the generalized JW 
transformations for the 3D case are interpreted as gauge transformations with 
the topological charge equal to 1. These transformations are more complicated 
than the transformations (1.11) from the paper \cite{23}.

Here we will consider generalized transformations  of the JW type for lattice 
systems from a different point of view. We will show that solutions  (1.10) 
and (1.11) of the equation (1.9) are not unique. More precisely, we will show 
that for the 2D case, as well as for the 3D case (and also in higher 
dimensions), there is a possibility to introduce two or more sets of Fermi 
creation and annihilation operators. Moreover, there is a nontrivial  
transposition algebra between various sets of Fermi operators. This fact was 
not realized by the authors of the papers \cite{20,21,22,23}, because of, as 
it seems to us, not a very clear and simple enough notation. Below we will 
point out  some possible applications of the generalized transformations  of 
the JW type in the form postulated by us to the analysis  of lattice models 
of statistical physics and in the graph theory, connected with the problem of 
calculation of generating functions  for self-avoiding walks (Hamiltonian  
cycles on a simple rectangular lattice , see \cite{26,27,28}). 

This problem is still under investigation, what could be seen e.g. from the 
recent paper of Gujrati \cite{24} devoted to a geometric description  of phase 
transitions in terms of diagrams and their growth functions. Moreover, as 
far as the author's knowledge is concerned, the multidimensional 
transformations of the JW type were not given in such a simple and convenient 
form, and their properties were not examined in the sense discussed above 
(algebra of transposition relations etc.). In what follows we adopt the 
notation accepted in contemporary literature \cite{12,Baxter,11,13,14} and we 
keep in touch with the spirit and ideas of the pioneering paper of Jordan and 
Wigner \cite{1}.

\section{The two-dimensional  transformation of Jordan-Wigner type}

Let us introduce three sets of $2^{\mbox{\eu nm}}$ dimensional Pauli matrices
$\tau^{x,y,z}_{nm}$    which can  be defined in the following way \cite{18},
$\{n(m)=1,2,\ldots =\mbox{\eu n(m)}\}$:
\begin{eqnarray}
\tau^x_{nm}=1\hspace{-1.1mm}\mbox{I}\bigotimes 1\hspace{-1.1mm}\mbox{I}\bigotimes 1\hspace{-1.1mm}\mbox{I}\ldots 1\hspace{-1.1mm}\mbox{I}\bigotimes\tau^x\bigotimes 1\hspace{-1.1mm}\mbox{I}\ldots
1\hspace{-1.1mm}\mbox{I}\bigotimes 1\hspace{-1.1mm}\mbox{I}\bigotimes 1\hspace{-1.1mm}\mbox{I}\hspace{2cm}(\mbox{{\eu nm} -- factors}) \nonumber\\
\tau^y_{nm}=1\hspace{-1.1mm}\mbox{I}\bigotimes 1\hspace{-1.1mm}\mbox{I}\bigotimes 1\hspace{-1.1mm}\mbox{I}\ldots 1\hspace{-1.1mm}\mbox{I}\bigotimes\tau^y\bigotimes 1\hspace{-1.1mm}\mbox{I}\ldots
1\hspace{-1.1mm}\mbox{I}\bigotimes 1\hspace{-1.1mm}\mbox{I}\bigotimes 1\hspace{-1.1mm}\mbox{I}\hspace{2cm}(\mbox{{\eu nm} -- factors}) \nonumber\\
\tau^z_{nm}=1\hspace{-1.1mm}\mbox{I}\bigotimes 1\hspace{-1.1mm}\mbox{I}\bigotimes 1\hspace{-1.1mm}\mbox{I}\ldots1\hspace{-1.1mm}\mbox{I}\bigotimes\tau^z\bigotimes 1\hspace{-1.1mm}\mbox{I}\ldots
1\hspace{-1.1mm}\mbox{I}\bigotimes 1\hspace{-1.1mm}\mbox{I}\bigotimes 1\hspace{-1.1mm}\mbox{I}\hspace{2cm}(\mbox{{\eu nm} -- factors})\label{7}
\end{eqnarray}
where the standard Pauli matrices $\tau^{x,y,z}$ are situated in these tensor 
products at the nm-th place (1\hspace{-1.2mm}I - the unit $2\times 2$ matrix). 
Further, the doubly indexed Pauli operators  $\tau^{\pm}_{nm}$ are defined in 
the way analogous, i.e. as
\begin{equation}
\tau^{\pm}_{nm}=\frac{1}{2}\left(\tau^z_{nm}\pm i\tau^y_{nm}\right),\;\;\;
\tau^x_{nm}=-2\left(\tau^+_{nm}\tau^-_{nm}-\frac{1}{2} \right),\;\;\;
\tau^z_{nm}=\tau^+_{nm}+\tau^-_{nm}\label{8}
\end{equation}
The Pauli operators satisfy anticommutation transposition relations for the 
same lattice node (nm):
\begin{equation}
\left\{\tau^+_{nm},\tau^-_{nm}\right\}_+=1, \;\;\;\;
\left(\tau^+_{nm}\right)^2=\left(\tau^-_{nm}\right)^2=0 ,
\label{9}
\end{equation}
and commutation relations for different nodes:
\begin{equation}
\left[\tau^\pm_{nm},\tau^\pm_{n'm'}\right]_-=0\hspace{2cm}
(nm)\neq (n'm').
\label{10}
\end{equation}

In other words the Pauli  operators (\ref{8}-\ref{10}) behave as Fermi 
operators for the same node and as Bose operators for different nodes. To 
accomplish the transition to the Fermi representation for  the whole 
lattice i.e. to the representation of Fermi creation and annihilation 
operators ($c^{\dag}_{nm},c_{nm}$) for the whole lattice, we introduce, in 
analogy to the one-dimensional J-W transformations (1.1), (1.4) two-
dimensional transformations of J-W type,  enabling us to express  the Fermi  
operators for a two-dimensional system ($c^{\dag}_{nm},c_{nm}$) by Pauli 
operators ($\tau^{\pm}_{nm}$). It occurs that in the two-dimensional case 
there exist two sets of such transformations, (we do not include here the 
inverse transformations of the type (\ref{6a})), which we write in the form:
\begin{eqnarray}
\alpha^{\dag}_{nm}&=&\exp\left( i\pi\sum^{n-1}_{k=1}\sum^{\mbox{\eu
m}}_{l=1}\tau^+_{kl} \tau^-_{kl}
+i\pi\sum^{m-1}_{l=1}\tau^+_{nl}\tau^-_{nl}\right)\tau^+_{nm}\nonumber,\\
\alpha_{nm}&=&\exp\left( i\pi\sum^{n-1}_{k=1}\sum^{\mbox{\eu
m}}_{l=1}\tau^+_{kl} \tau^-_{kl}
+i\pi\sum^{m-1}_{l=1}\tau^+_{nl}\tau^-_{nl}\right)\tau^-_{nm},
\label{11}\\
\beta^{\dag}_{nm}&=&\exp\left( i\pi\sum^{\mbox{\eu n}}_{k=1}\sum^{m-1}
_{l=1}\tau^+_{kl} \tau^-_{kl}
+i\pi\sum^{n-1}_{k=1}\tau^+_{km}\tau^-_{km}\right)\tau^+_{nm}\nonumber,\\
\beta_{nm}&=&\exp\left( i\pi\sum^{\mbox{\eu n}}_{k=1}\sum^{m-1}_{l=1}\tau^+_{kl} \tau^-_{kl}
+i\pi\sum^{n-1}_{k=1}\tau^+_{km}\tau^-_{km}\right)\tau^-_{nm}.
\label{12}
\end{eqnarray}

It is straightforward to check, using the relations (\ref{9}) and (\ref{10})
that the operators 
$(\alpha^{\dag}_{nm},\alpha_{nm})$ and $(\beta^{\dag}_{nm},\beta_{nm})$ are 
actually Fermi operators for all the lattice i.e. that they satisfy 
anticommutation transposition relations for all nodes:
\begin{eqnarray}
\{\alpha^{\dag}_{nm},\alpha_{nm}\}_+=1,\;\;\;\;
(\alpha^{\dag}_{nm})^2=(\alpha_{nm})^2=0,\\
\{\alpha^{\dag}_{nm},\alpha^{\dag}_{n'm'}\}_+=\{\alpha^{\dag}_{nm},\alpha_
{n'm'}\}_+=\{\alpha_{nm},\alpha_{n'm'}\}_+=0\hspace{1cm}(nm)\neq(n'm'),\nonumber
\end{eqnarray}
and analogously for  $\beta$-operators. The inverse transformations to
(\ref{11}-\ref{12}) are:
\begin{eqnarray}
\tau^+_{nm}&=&\exp\left( i\pi\sum^{n-1}_{k=1}\sum^{\mbox{\eu
m}}_{l=1}\alpha^{\dag}_{kl} \alpha_{kl}
+i\pi\sum^{m-1}_{l=1}\alpha^{\dag}_{nl}\alpha_{nl}\right)\alpha^{\dag}_{nm}\nonumber,\\
\tau_{nm}^+&=&\exp\left( i\pi\sum^{\mbox{\eu n}}_{k=1}\sum^{
m-1}_{l=1}\beta^{\dag}_{kl} \beta_{kl}
+i\pi\sum^{n-1}_{k=1}\beta^{\dag}_{km}\beta_{km}\right)\beta^{\dag}_{nm},
\label{14}
\end{eqnarray}
and analogously for  $\tau^-_{nm}$.

The checking procedure could be easily performed provided we take into
consideration the following relations:
\begin{equation}
\exp{\left(
i\pi\sum_{nm}\tau^+_{nm}\tau^-_{nm}\right)}=\prod_{nm}(1-2\tau^+_{nm}\tau_{nm})
=\prod_{nm}(\tau^x_{nm})\label{15}
\end{equation}
That's obvious:
\begin{equation}
\tau^+_{nm}\tau^-_{nm}=\alpha^{dag}_{nm}\alpha_{nm}, \;\;\;
\tau^+_{nm}\tau^-_{nm}=\beta^{\dag}_{nm}\beta_{nm}, \;\;\;
\alpha^{\dag}_{nm}\alpha_{nm}=\beta^{\dag}_{nm}\beta_{nm}\label{16}
\end{equation}
The last formula in (\ref{16}) expresses the condition of local equality of
occupation numbers for $\alpha$- and $\beta$-fermions in the same node.

It is a simple matter to see that in the discrete case the solution (1.10) 
can be identified with the first pair of transformations (2.8), if we 
introduce the following correspondence:
\begin{equation}
x_1\to n,\hspace{1cm} z_1\to k;\hspace{1cm} x_2\to m,\hspace{1cm}z_2\to l
\end{equation}
Afterwards, we have 
$$ \sum_{\bf Z}w({\bf x,z})\alpha^{\dag}({\bf z})\alpha ({\bf z}) \to
\sum^{n-1}_{k=1}\sum^{M}_{l=1}\alpha^{\dag}_{kl}\alpha_{kl}+\sum^{m-1}_{l=1}
\alpha^{\dag}_{nl}\alpha_{nl} $$
Analogously, for
\begin{equation}
x_1\to m,\hspace{1cm} z_1\to l;\hspace{1cm} x_2\to n,\hspace{1cm}z_2\to k , 
\end{equation}
we obtain the second pair of transformations (2.8). From (2.11-2.12) it 
follows that the pair of transformations (2.8) can be written in the form 
(in the notation from \cite{23}):
\begin{equation}
w({\bf x,z})=\Theta(x_i-z_i)(1-\delta_{x_iz_i})+\Theta(x_j-z_j)\delta_
{x_iz_i},\hspace{1cm}(i\neq j)=1,2.
\end{equation}
From (2.13) drop out two others inverse transformations, analogous to 
transformations (1.5) in the one-dimensional case. Obviously, (2.11-2.12) 
corresponded to the symmetric transposition group $S_2$ and, therefore, to 
the complete set of transformations  for  the 2D case in the  discrete case 
corresponds the group $S_2$. This set of transformations should be 
complemented by a group of inverse transformations, analogous to 
transformations (1.5) in one-dimensional case. These inverse transformations 
could always be written out if necessary. 

It follows from (\ref{11}--\ref{12}), (\ref{14}) and (\ref{16})  that 
operators $(\alpha^{\dag}_{nm},\alpha_{nm})$ and $(\beta^{\dag}_{nm},\beta_
{nm})$ are connected by relations of the form:
\begin{eqnarray}
\alpha^{\dag}_{nm}=\exp(i\pi \phi_{nm})\beta^{\dag}_{nm}, \;\;\;\;
\alpha_{nm}=\exp(i\pi \phi_{nm})\beta_{nm}\nonumber\\
\phi_{nm}=\left[ \sum^{\mbox{\eu n}}_{k=n+1}\sum^{m-1}_{l=1}
+\sum^{n-1}_{k=1}\sum^{\mbox{\eu m}}_{l=m+1}\right]
\alpha^{\dag}_{kl}\alpha_{kl}=\left[\ldots\right]\beta^{\dag}_{kl}\beta_{kl}
\label{17}
\end{eqnarray}
It is obvious that the operators $\phi_{nm}$ commute with operators 
$(\alpha^{\dag}_{nm},\alpha_{nm})$ and $(\beta^{\dag}_{n'm'},\beta_{n'm'})$ 
in the same node:
\begin{equation}
[\phi_{nm},\alpha^{\dag}_{nm}]_-=\ldots =\ldots =[\phi_{nm},\beta_{nm}]_-=0,
\end{equation}
because the occupation numbers with the index $(nm)$ drop out of the 
$\phi_{nm}$.

It is easy to see that in the one-dimensional case the transformations 
introduced  become identical to the one-dimensional  J-W 
transformations (1.1), (1.4).  We should stress here that  an inverse 
transition does not exist, i.e. a transformation from  the one-dimensional 
J-W transitions (1.1), (1.4) to their two-dimensional analogue (\ref{11}-
\ref{12}). In other words a "derivation" of the two-dimensional transformations 
(\ref{11}-\ref{12}) from the one-dimensional J-W transformations  (1.1), 
(1.4)  is not possible using, for example, the lexicological order \cite{19} 
for the current double indexed variables  $\tau^\pm_{nm}$, or using different 
types of order. No doubts, we obtain  in this case the well known  one-
dimensional transformations, not more. Of course, we use extensively the 
ideas of Jordan and Wigner to obtain the multidimensional analogue of the 
transformations.

Now, let us find the transposition relations for the operators 
$(\alpha^{\dag}_{nm},\alpha_{nm})$ and $(\beta^{\dag}_{n'm'},\beta_{n'm'})$. 
Firstly, using the relations:
\begin{equation}
\exp(i\pi\alpha^{\dag}_{nm}\alpha_{nm})=(1-2\alpha^{\dag}_{nm}\alpha_{nm})=
(-1)^{\alpha^{\dag}_{nm}\alpha_{nm}}, 
\end{equation}
it is easy to find the following transposition relations:
\begin{equation}
\{(-1)^{\alpha^{\dag}_{nm}\alpha_{nm}},\alpha_{nm}\}_+=
\{(-1)^{\alpha^{\dag}_{nm}\alpha_{nm}},\alpha^{\dag}_{nm}\}_+=
\{(-1)^{\alpha^{\dag}_{nm}\alpha_{nm}},\beta_{nm}\}_+=
\{(-1)^{\alpha^{\dag}_{nm}\alpha_{nm}},\beta^{\dag}_{nm}\}_+=0
\label{20}
\end{equation}
where the equality of occupation numbers for the $\alpha$ and $\beta$ fermions
(\ref{16}) has been used. The straightforward calculation gives the following
transposition relations:
\begin{eqnarray}
\{\alpha^{\dag}_{nm},\beta_{nm}\}_+\!\!&\!\!=\!\!&\!\!
\{\beta^{\dag}_{nm},\alpha_{nm}\}_+=(-1)^{\phi_{nm}}\label{21}\\ 
\left[ \alpha_{nm},\beta_{n'm'}  \right] _{-} \!\!&\!\!=\!\!&\!\!
\ldots =
\left[ \alpha^{\dag}_{nm},\beta^{\dag}_{n'm'} \right] _{-}=0,\mbox{for}
\left(\begin{array}{c c}
n'\leq n-1,&m'\geq m+1\\
n'\geq n+1,&m'\leq m-1
\end{array}\right)\nonumber\\ & & \\
\{\alpha_{nm},\beta_{n'm'}\}_{+}\!\!&\!\!=\!\!&\!\!
\{\alpha^{\dag}_{nm},\beta^{\dag}_{n'm'}\}_{+}=0\label{23}
\end{eqnarray}
for all the cases, where the operators $\phi_{nm}$ in (\ref{21}) are defined 
by the formula (\ref{17})  and we use the equality:
$$\exp (i\pi\phi_{nm})=(-1)^{\phi_{nm}}.$$
The transformation relations (\ref{21}-\ref{23}) are simply illustrated in 
the Fig.1, where the distinguished operator $\alpha_{nm}$  for a fixed node
$(nm)$ commutes with $\beta$-operators for the nodes $(n',m')$, denoted  by a 
cross $(\times )$. For all other nodes $\alpha$- and $\beta$-operators 
anticommute. On the other hand, we can easily obtain the expression for the 
commutation for the same node $(nm)$
\begin{equation}
\alpha_{nm}\beta^{\dag}_{nm}-\beta^{\dag}_{nm}\alpha_{nm}=(-1)^{\beta^{\dag}_
{nm}\beta_{nm}}(-1)^{\phi_{nm}}
\end{equation}
In this way we obtain a rather specific structure of transposition relations 
for $\alpha$- and $\beta$-operators for the lattice, in spite of the 
fact that this structure has some symmetry. 

Here is the proper place to make a small digression and compare the situation
described above with the situation in which we  use the method of second
quantization. In the latter case for a system constituting of different 
particles the operators of  second quantization are introduced. The operators 
that are assigned to bosons and fermions commute. On the other hand, for the
operators  assigned  to different fermions it is usually postulated without
proof \cite{13} that in the framework of the non-relativistic theory these
operators could be formally treated either as commuting or as anticommuting. 
For both possible assumptions about the transposition relations the 
application of the method of second  quantization gives the same result. On 
the other hand, as far as the relativistic theory is concerned, where some 
transmutations of particles  are possible, we should treat the creation and 
annihilation operators for different fermions as anticommuting.

In the case we deal with, we operate formally with "quasiparticles" of the 
$\alpha$- and $\beta$-type, which, treated separately, are subjected to the 
Fermi statistics. In contrast, the transposition relations among the members 
of these two sets of operators depend on the relative position of the 
"quasiparticles" in the nodes of the lattice. As far as it is known to the 
author, such a situation has not occured in  quantum physics. 

The fact that in the two-dimensional case there are two sets of  
transformations (neglecting the inversional transformations of the  type 
(\ref{6a}), discussed below) of the JW type (\ref{11}) and (\ref{12}) is, in 
a way, justified if we consider statistical mechanics of two- and three-
dimensional lattice models with the nearest-neighbours interactions. Namely, 
assume we managed for one of such models (for example, for the Ising model or 
another model describing two level states of any system) to express the 
Hamiltonian in terms of the  doubly indexed Pauli operators $\tau^{\pm}_{nm}$ 
and the components of the Hamiltonian are of the form: 
\begin{equation}
\tau^+_{nm}\tau^+_{n+1,m},\hspace{1cm}
\tau^+_{nm}\tau^-_{n,m+1},\hspace{.5cm}etc. \label{25}
\end{equation}

Then it is easy to obtain for the first component of  (\ref{25}), after 
application of the transformations (\ref{12}), the expression:
\begin{equation}
\tau^+_{nm}\tau^+_{n+1,m}=\beta^{\dag}_{nm}(1-2\beta^{\dag}_{nm}\beta_{nm})
\beta^{\dag}_{n+1,m}=\beta^{\dag}_{nm}\beta^{\dag}_{n+1,m} ,
\end{equation}
and for the second component of (\ref{25}), after application of 
transformations (\ref{11}), the expression:
\begin{equation}
\tau^+_{nm}\tau^-_{n,m+1}=\alpha^{\dag}_{nm}(1-2\alpha^{\dag}_{nm}\alpha_{nm})
\alpha_{n,m+1}=\alpha^{\dag}_{nm}\alpha_{n,m+1},
\end{equation}
On the other hand, if we apply to the first component of (\ref{25}) the 
transformations (\ref{11}),  we obtain:
\begin{eqnarray}
\tau^+_{nm}\tau^+_{n+1,m}&=&(-1)^{\phi_{nm}}\alpha^{\dag}_{nm}(-1)^{\phi_
{n+1,m}}\alpha^{\dag}_{n+1,m}=(-1)^{\chi_{nm}}\alpha^{\dag}_{nm}\alpha^{\dag}
_{n+1,m},\nonumber\\
\chi_{nm}&=&\sum^{\mbox{\eu
m}}_{l=m+1}\alpha^{\dag}_{nl}\alpha_{nl}+\sum^{m-1}_{l=1}\alpha^{\dag}_{n+1,l}
\alpha_{n+1,l}
\end{eqnarray}
and if we apply to the second component of (\ref{25}) the transformations 
(\ref{12}), we obtain:
\begin{eqnarray}
\tau^+_{nm}\tau^-_{n,m+1}&=&
(-1)^{\phi_{nm}}\beta^{\dag}_{nm}(-1)^{\phi_{n,m+1}}\beta_{n,m+1}=
(-1)^{\rho_{nm}}\beta^{\dag}_{nm}\beta_{n,m+1},\nonumber\\
\rho_{nm}&=&
\sum^{\mbox{\eu n}}_{k=n+1}\beta^{\dag}_{km}\beta_{km}+
\sum^{n-1}_{k=1}\beta^{\dag}_{k,m+1}\beta_{k,m+1}\label{29}
\end{eqnarray}
In this way, in the expressions (2.25) and (\ref{29}) some unpleasant phase 
factors $(-1)^{\chi_{nm}}$  and  $(-1)^{\rho_{nm}}$ are contained.  
It is this place where certain difficulties appear, connected with the  
attempts to apply the JW transformations (2.8) in the solution of the 3D 
Ising model using the approach introduced in the paper \cite{11}.  In some 
cases, for example during the calculation of energy of the ground state, 
those phase factors could be eliminated from considerations, because:
\begin{equation}
(-1)^{\chi_{nm}}|0\rangle=1|0\rangle, \;\;\;\;\;
(-1)^{\rho_{nm}}|0\rangle=1|0\rangle
\end{equation}
where: $|0\rangle$ - the vacuum state; 
$\alpha_{nm}|0\rangle=\beta_{nm}|0\rangle=0$. As we see, application of the 
transformations (\ref{12}) to the first index (n) and the transformations 
(\ref{11}) to the second index $(m)$ does not lead to occurrence of these 
phase factors. As a result, this fact implies the possibility of   
diagonalization by transformation to the momentum representation. 
Unfortunately, there occur some other obstacles for the diagonalization, 
which will be discussed  elsewhere.

The arguments given above for existence of at least two sets of 
transformations (\ref{11}) and (\ref{12}) of the J-W type in the two-dimensional 
case are, of course not, rigorous and are presented here only as some 
guiding devices. Possible future physical interpretation of these results has 
nothing to do with the mathematical fact of existence of two nontrivial 
transformations for two-dimensional systems. Introduction of inverse 
transformations of the type (\ref{6a}) for the two-dimensional case do not 
lead to any new transformations, and do not change the symmetry of the 
transposition relations. As it is seen below, in the three-dimensional case
the situation is much more complicated and the arguments given above are, in 
general, powerless.

In the papers \cite{25,26,27,28} of the author was shown a nontrivial example 
of application of a pair of the JW transformations (2.8) to the problem of 
deriving the generating function for Hamiltonian cycles on a simple 
rectangular  lattice with $N\times M$ nodes. One of the key moments in the 
papers is simultaneous application of the pair of JW transformations (2.8), 
and not of only of one of them. We will investigate in detail application of 
the pair (2.8) to one of the possible solutions of the 2D Ising-Onsager 
problem in the external field \cite{26}.

\section{The three-dimensional transformation of the J-W type}

In the  three-dimensional case we introduce three sets of $2^{\mbox{\eu nmk}}$
-dimensional Pauli matrices $\tau^{x,y,z}_{nmk}$ ($n=1,2,\ldots,${\eu n}; 
$m=1,2,\ldots,${\eu m}; $k=1,2,\ldots,${\eu k}), which are defined analogously 
to the two-dimensional case (\ref{7}). Further we introduce three-index 
Pauli operators $\tau^{\pm}_{nmk}$ by formulae:
\begin{equation}
\tau^{\pm}_{nmk}=2^{-1}(\tau^z_{nmk}\pm i\tau^y_{nmk})
\end{equation}
which satisfy anticommutation transposition relations for the same lattice 
node $(nmk)$:
\begin{equation}
\left\{\tau^+_{nmk},\tau^-_{nmk}\right\}_+=1,\;\;\;\;
\left(\tau^+_{nmk}\right)^2=\left(\tau^-_{nmk}\right)^2=0\label{32}
\end{equation}
and commutation relations for different lattice nodes:
\begin{equation}
\left[\tau^\pm_{nmk},\tau^\pm_{n'm'k'}\right]_-=0\hspace{2cm}
(nmk)\neq (n'm'k').                              \label{33}
\end{equation}
It occurs that in the three-dimensional case there exist six (not including 
the inverse transformations of the type \ref{6a}) sets of 
transformations of the J-W  type, which could be represented in the form:
\begin{eqnarray}
\tau^+_{nmk}\!\!&\!\!=\!\!&\!\!\exp \left[ i\pi\left(\sum^{\mbox{\eu n}}_{s=1}
\sum^{\mbox{\eu
m}}_{p=1}\sum^{k-1}_{q=1}\alpha^{\dag}_{spq}\alpha_{spq}+
\sum^{\mbox{\eu n}}_{s=1}\sum^{m-1}_{p=1}\alpha^{\dag}_{spk}\alpha_{spk}+
\sum^{n-1}_{s=1}\alpha^{\dag}_{smk}\alpha_{smk}\right)\right]\alpha^{\dag}_{nmk}
\nonumber\\\label{34}\\
\tau^+_{nmk}\!\!&\!\!=\!\!&\!\!\exp \left[ i\pi\left(\sum^{\mbox{\eu n}}_{s=1}
\sum^{\mbox{\eu
m}}_{p=1}\sum^{k-1}_{q=1}\beta^{\dag}_{spq}\beta_{spq}+
\sum^{ n-1}_{s=1}\sum^{\mbox{\eu m}}_{p=1}\beta^{\dag}_{spk}\beta_{spk}+
\sum^{m-1}_{p=1}\beta^{\dag}_{npk}\beta_{npk}\right)\right]\beta^{\dag}_{nmk}
\nonumber\\\label{35}\\
\tau^+_{nmk}\!\!&\!\!=\!\!&\!\!\exp \left[ i\pi\left(\sum^{\mbox{\eu n}}_{s=1}
\sum^{m-1}_{p=1}\sum^{\mbox{\eu k}}_{q=1}\gamma^{\dag}_{spq}\gamma_{spq}+
\sum^{\mbox{\eu n}}_{s=1}\sum^{k-1}_{q=1}\gamma^{\dag}_{smq}\gamma_{smq}+
\sum^{n-1}_{s=1}\gamma^{\dag}_{smk}\gamma_{smk}\right)\right]\gamma^{\dag}_{nmk}
\nonumber\\\label{36}\\
\tau^+_{nmk}\!\!&\!\!=\!\!&\!\!\exp \left[ i\pi\left(\sum^{\mbox{\eu n}}_{s=1}
\sum^{m-1}_{p=1}\sum^{\mbox{\eu k}}_{q=1}\eta^{\dag}_{spq}\eta_{spq}+
\sum^{n-1}_{s=1}\sum^{\mbox{\eu k}}_{q=1}\eta^{\dag}_{smq}\eta_{smq}+
\sum^{k-1}_{q=1}\eta^{\dag}_{nmq}\eta_{nmq}\right)\right]\eta^{\dag}_{nmk}
\nonumber\\\label{37}\\
\tau^+_{nmk}\!\!&\!\!=\!\!&\!\!\exp \left[ i\pi\left(\sum^{n-1}_{s=1}
\sum^{\mbox{\eu m}}_{p=1}\sum^{\mbox{\eu k}}_{q=1}\omega^{\dag}_{spq}\omega_{spq}+
\sum^{\mbox{\eu m}}_{p=1}\sum^{k-1}_{q=1}\omega^{\dag}_{npq}\omega_{npq}+
\sum^{m-1}_{p=1}\omega^{\dag}_{npk}\omega_{npk}\right)\right]\omega^{\dag}_{nmk}
\nonumber\\\label{38}\\
\tau^+_{nmk}\!\!&\!\!=\!\!&\!\!\exp \left[ i\pi\left(\sum^{n-1}_{s=1}
\sum^{\mbox{\eu m}}_{p=1}\sum^{\mbox{\eu k}}_{q=1}\theta^{\dag}_{spq}\theta_{spq}+
\sum^{m-1}_{p=1}\sum^{\mbox{\eu k}}_{q=1}\theta^{\dag}_{npq}\theta_{npq}+
\sum^{k-1}_{q=1}\theta^{\dag}_{nmq}\theta_{nmq}\right)\right]\theta^{\dag}_{nmk}
\nonumber\\\label{39}
\end{eqnarray}
and  analogously for the operators $\tau^-_{nmk}$. For the sake of 
completeness of the exposition, we have written here six transformations, 
which enable us to express the Pauli operators  $\tau^{\pm}_{nmk}$ by the 
Fermi creation and annihilation operators $(\alpha^{\dag}_{nmk},\alpha_{nmk},
\ldots,\theta_{nmk})$. Applying the formulae   (\ref{32}),  (\ref{33}) and the 
formulae of the type (\ref{20}), it is easy to show that the operators
$(\alpha^{\dag}_{nmk},\alpha_{nmk},\ldots ,\theta_{nmk})$  satisfy the 
anticommutation transposition relations:
\begin{eqnarray}
\{\alpha^{\dag}_{nmk},\alpha_{nmk}\}_+= 1, \;\;\;\;
(\alpha^{\dag}_{nmk})^2=(\alpha_{nmk})^2=0\nonumber\\
\{\alpha^{\dag}_{nmk},\alpha^{\dag}_{n'm'k' }\}_+=\ldots =
\{\alpha_{nmk},\alpha_{n'm'k'}\}_+=0\hspace{.3cm}(nmk)\neq (n'm'k')
\end{eqnarray}
etc., as could be straightforwardly checked. There exist also inverse
transformations:
\begin{equation}
\alpha^{\dag}_{nmk}=\exp \left[ i\pi\left(\sum^{\mbox{\eu n}}_{s=1}
\sum^{\mbox{\eu m}}_{p=1}\sum^{k-1}_{q=1}\tau^{+}_{spq}\tau^-_{spq}+
\sum^{\mbox{\eu n}}_{s=1}\sum^{m-1}_{p=1}\tau^+_{spk}\tau^-_{spk}+
\sum^{n-1}_{s=1}\tau^+_{smk}\tau^-_{smk}\right)\right]\tau^{+}_{nmk}
\end{equation}
etc., from which we can obtain easily the equations:
\begin{eqnarray}
\tau^+_{nmk}\tau^-_{nmk}=\alpha^{\dag}_{nmk}\alpha_{nmk}=\beta^{\dag}_{nmk}
\beta_{nmk}=\gamma^{\dag}_{nmk}\gamma_{nmk}=\eta^{\dag}_{nmk}\eta_{nmk}=
\nonumber\\
=\omega^{\dag}_{nmk}\omega_{nmk}=\theta^{\dag}_{nmk}\theta_{nmk}\label{42}
\end{eqnarray}
using relations of the type (\ref{15}), written for the three-dimensional 
case. The relations (\ref{42}) express the conditions of equality of local occupation
numbers for $\alpha$- ,$\beta$- , $\gamma$- , $\eta$- , $\omega$- and 
$\theta$- fermions for the same lattice node $(nmk)$. In analogy  to the two-
dimensional case, the operators
$(\alpha^{\dag}_{nmk},\ldots ,\theta_{nmk})$ are connected by canonical 
nonlinear transformations:
\begin{eqnarray}
\alpha^{\dag}_{nmk}&=&(-1)^{\phi_{nmk}}\beta^{\dag}_{nmk}\nonumber\\
\alpha_{nmk}&=&(-1)^{\phi_{nmk}}\beta_{nmk}\nonumber\\
\phi_{nmk}&=&\left[\sum^{\mbox{\eu n}}_{s=n+1}\sum^{m-1}_{p=1}+
 \sum^{n-1}_{s=1}\sum^{\mbox{\eu m}}_{p=m+1}\right]\alpha^{\dag}_{spk}\alpha_{spk}
\label{43};\\
\alpha^{\dag}_{nmk}&=&(-1)^{\psi_{nmk}}\gamma^{\dag}_{nmk}\nonumber\\
\alpha_{nmk}&=&(-1)^{\psi_{nmk}}\gamma_{nmk}\nonumber\\
\psi_{nmk}&=&\sum^{\mbox{\eu n}}_{s=1}\left[\sum^{\mbox{\eu m}}_{p=m+1}
\sum^{k-1}_{q=1}+
 \sum^{m-1}_{p=1}\sum^{\mbox{\eu k}}_{q=k+1}\right]\alpha^{\dag}_{spq}\alpha_{spq}
\label{44};\\
\beta^{\dag}_{nmk}&=&(-1)^{\chi_{nmk}}\gamma^{\dag}_{nmk}\nonumber\\
\beta_{nmk}&=&(-1)^{\chi_{nmk}}\gamma_{nmk}\nonumber\\
\chi_{nmk}&=&\sum^{\mbox{\eu n}}_{s=1}\left[\sum^{\mbox{\eu m}}_{p=m+1}
\sum^{k-1}_{q=1}+
 \sum^{m-1}_{p=1}\sum^{\mbox{\eu k}}_{q=k}\right]\beta^{\dag}_{spq}\beta_{spq}
+\sum^{n-1}_{s=1}\sum^{\mbox{\eu
m}}_{p=1}\beta^{\dag}_{spk}\beta_{spk}\nonumber\nopagebreak\\\nopagebreak
& &+\sum^{m-1}_{p=1}\beta^{\dag}_{npk}\beta_{npk}+
\sum^{n-1}_{s=1}\beta^{\dag}_{smk}\beta_{smk};
\end{eqnarray}
and 12 pairs of further transformations, which could be easily written, if 
necessary. The operators $\phi_{nmk}$, $\psi_{nmk}$ etc. obviously commute 
with the operators $(\alpha^{\dag}_{nmk},\ldots ,\theta_{nmk})$ in the same 
node, because of lack in the operators  $\phi_{nmk}$, $\psi_{nmk}$ etc. of the 
operators of occupation numbers indexed by $(nmk)$. It is also a rather easy 
task to prove that the transformations of the J-W type introduced 
above by the formulae (\ref{34}-\ref{39}) for the three-dimensional case 
reduce to transformations (\ref{11}), (\ref{12}) in the two-dimensional case.

Similarly as in the 2D case, correspondence  between transformations (3.4-3.9) 
and the solution (12) can be established in the three-dimensional case on the 
basis of the symmetric group $S_3$. Indeed, the transformations (3.4-3.9) in 
the notation from the paper \cite{23} can be written in the form:
\begin{eqnarray}
 w({\bf x,z})&=&\Theta(x_i-z_i)(1-\delta_{x_iz_i})+\Theta(x_j-z_j)\delta_{x_iz_i}
 (1-\delta_{x_jz_j})\nonumber\\
 &+&\Theta(x_k-z_k)\delta_{x_iz_i}\delta_{x_jz_j},\hspace{1cm}\nonumber\\
(i\neq j\neq k)&=&1,2,3.
\end{eqnarray}
It should be noticed that from (3.15) drop out the inverse transformations of 
the type (1.5), generalized to the 3D case. Obviously in the d-dimensional 
case the complete number of  transformations of the   JW type is equal to d! 
(neglecting the inverse transformations), and the correspondence can be 
established using the symmetric group $S_d$.

There are no principal obstacles against full analysis of transposition 
relations for the operators $(\alpha^{\dag}_{nmk},\ldots ,\theta_{nmk})$ 
and, if necessary, we can write all the relations we want. Here we consider 
the transposition relations only for the operators 
$(\alpha^{\dag}_{nmk},\beta^{\dag}_{n'm'k'},\gamma^+_{n''m''k''},)$, 
to get feeling of the  "geometric structure" of these relations for the three-
dimensional case. First of all let us observe that, according to (\ref{43}), 
transposition relations for  $(\alpha^{\dag}_{nmk},\alpha_{nmk})$ 
and $(\beta^{\dag}_{n'm'k'},\beta_{n'm'k'})$ for $k=k'$  are of the form 
(\ref{21}-\ref{23}), where it is sufficient to add the third index $k$ to all 
the operators. In other words, transposition relations in the plane 
($nm/k=const$) for $\alpha$- and $\beta$- operators behave like in the two-
dimensional case. This fact could be expected, because for the fixed $k$ we 
deal actually with two-dimensional transformations of the J-W type (\ref{11}
-\ref{12}). It is easy to see that for all of the three mutually orthogonal 
planes there exists a pair of operators, for which the transformation 
relations are of the form (\ref{21}-\ref{23}). In accordance with (\ref{34}-
\ref{39}) for the plane $(mk/n=const)$ such a pair will be the pair of 
operators $\omega-\theta $, and for the plane $(nk/m=const)$ - the pair of 
operators $\gamma-\eta$. Further, according to (\ref{43}) for  $(k\neq k')$, 
$\alpha$- and $\beta$ - operators anticommute for any $(nm)$ i.e.
\begin{equation}
\{\alpha^{\dag}_{nmk},\beta^{\dag}_{n'm'k'}\}_+=\ldots =
\{\alpha_{nmk},\beta_{n'm'k'}\}_+=0,\hspace{1cm} (k\neq k')
\label{46}
\end{equation}
It is obvious that  also the pairs of operators $\omega - \theta$ 
anticommute for  $(n\neq n')$ and the pair of operators $\gamma - \eta$ 
anticommutes for $(m\neq m')$. Now, from (\ref{44}) there follow 
transposition relations for $\alpha$- and $\gamma$-operators of the form 
analogous to (\ref{21}-\ref{23}) i.e.:
\begin{eqnarray}
[\alpha_{nmk},\gamma_{n'm'k'}]_-&=&\ldots =\ldots =0,\hspace{1cm} \mbox{for}\left(
\begin{array}{c c c}m'\leq m-1&,&k'\geq k+1\\
m'\geq m+1&,&k'\leq k-1
\end{array}\right),\nonumber\\
\{\alpha_{nmk},\gamma_{n'm'k'}\}_+&=&\ldots
=\{\alpha^{\dag}_{nmk},\gamma_{n'm'k'}\}_+=0,
\label{47}
\end{eqnarray}
in all other cases, with the only difference that the equations (\ref{47}) are 
satisfied for any $n$ and $n'$. Analogously, transposition relations for 
other pairs of operators are considered. In the general case the symmetry 
characteristic for the two-dimensional case (see (\ref{21} --\ref{23}) and 
Fig.1) disappears.

Therefore, in the case of the three-dimensional space there exist six 
nontrivial transformations of the J-W type (\ref{34}-\ref{39}) and the 
algebra of their transposition relations is much more complicated than the  
analogous algebra in the two-dimensional case. Some examples of application 
of the transposition relations (\ref{34}-\ref{39}) will be considered 
elsewhere, where the three-dimensional Ising model with and without external 
magnetic field is considered from the new point of view, as well as other 
models of statistical mechanics and physics etc. \cite{Baxter,25,26,27}.

We believe it is reasonable to notice here one beautiful fact connected with
generalized transformations of the JW type for the $d\geq 3 $ case. Namely, 
for lattice models with nearest neighbours interactions (for example, for the 
Ising model), the statistical sum of the system  can be represented in  terms 
of three-index Pauli operators $\tau^\pm\,_{nmk}$, which will enter the sum as 
bi-linear products of the type:
\begin{equation}
\tau^+_{nmk}\tau^+_{n+1,mk},
\end{equation}
etc., see \cite{27,28}. Then one can easily realize that among 
transformations (3.4)-(3.9) there are two, (3.4) and (3.5), which after 
application of which to (3.18) lead to the expressions 
$\alpha^{\dag}_{nmk}\alpha^{\dag}_{n+1,mk}$ and $\gamma^{\dag}_{nmk}
\gamma^{\dag}_{n+1,mk}$ in which the phase factors of the type 
$(-1)^{\chi_{nmk}}$ (2.25) etc. are not present, and the same applies to 
indices $m$ and $k$. In other words, there exist two equivalent (in the sense 
of absence of the phase factors) transformations of the JW type for each 
degree of freedom of Pauli variables. This is  a sort of degeneracy in  each 
index. For the 3D case the number $N_d$ of generalized transformations of the 
JW type is equal to $N_d=3!=6$, and the "degree" of degeneracy in every index 
is equal to $v=(N_d/d)=2$. Then in the general case we have $v=(N_d/d)=(d!/d)$ 
and for  $d=2$, $v=1$ as we have mentioned above (2.22)-(2.26). 

\section{Conclusions}

We believe we have menaged to show in  this paper advantages and simplicity 
of the generalized transformations of the JW type introduced  above. 
Especially important is the fact that this formulation of the JW 
transformations enabled us to find the entire sequence of sets of the 
transformations, and to investigate the algebra of transposition relations 
for various sets of Fermi operators. Moreover, as far as the  analysis of 
topological aspects of the transformations and consideration of their 
continuous counterparts are concerned, it seems 
that the notation from the papers \cite{20}-\cite{23} is more convenient. 

We omitted in this paper consideration of the problem of correspondence 
between discrete generalized transformations of the JW type, introduced in 
this paper and analogous transformations given in the papers \cite{20}-
\cite{23}, which are their continuous counterparts thereof. The reason is 
there are still many unclear points that need some analysis in future papers. 
Especially interesting, both from the physical (in the framework of possible 
applications) and mathematical point of view would be examination of the 
connection between our transformations and transformations taken from the 
paper \cite{22}. Such analysis  is missing also in the paper \cite{23}. This 
statement applies to the discrete case, but to the continuous one as well 
provided there exists a formal way to take the continuous limit for the 
lattice constant {\bf a}$\to 0$. For example, yet in the 2D case such a 
formal transition to the continuous limit could result in singularities of 
the section type along lines $x=const$ and $y=const$, where $\alpha (x,y)$, 
$\beta (x,y)$- densities in a fixed point (x,y) in the chosen coordinate
system ({\bf e}$_1$,{\bf e}$_2$) (relative to the transposition relations 
between $\alpha$- and $\beta$-operators (2.18)-(2.21). Here also some other 
problems appear which we will be explored in future papers.

The attention of physicists and mathematicians was and still is in the field 
theory and in connections there of with various models of classical and 
quantum statistical mechanics (see for example, \cite{12}, \cite{Baxter}, 
\cite{Gaudin}) and the literature cited in these papers). It is known 
\cite{12},\cite{Baxter}, \cite{Gaudin} that, in some cases a deep connection 
between the models of quantum field theory and the models of statistical 
mechanics has been discovered.

We hope that in the given form the proposed above JW type transformations for 
the 2D and 3D could be a valuable tool in the analysis of the 
already known models of statistical mechanics and quantum field theory, as 
well as they are expected to initiate formulation of new problems in these 
areas of theoretical physics. With the help of generalized J-W transformations 
type a new approach for Lenz-Ising-Onsager problem (LIO) has been made 
\cite{25,26,27,28,29}. For example, in the frame of this approach 2D LIO 
problem in the asymptotic magnetic field has been solved \cite{26}: 
\begin{eqnarray}
-\beta f_2(h\rightarrow 0)\sim\ln 2+2\ln(\cosh h/2)+ \nonumber \\
\frac{1}{2\pi^2}\int^{\pi}_0\int^{\pi}_0\ln[
\cosh{2K_1^*}\cosh{2K_2^*}-\sinh{2K_1^*}\cos q-\sinh{2K_2^*}\cos p]dq dp,
\end{eqnarray}
where the parameters 
$(K_{1,2},h)$ are to be renormalised in the following way $(K_{1,2}\geq 0)$:
\begin{eqnarray}
\sinh2K^*_{1,2}=\beta_{1,2}[\sinh2K_{1,2}(1-\tanh^2(h/2)],\nonumber \\
\cosh(2K^*_{1,2})=\beta_{1,2}[\cosh2K_{1,2}+\tanh^2(h/2)\sinh2K_{1,2}],
\nonumber \\
\beta_{1,2}=[1+2\tanh^2(h/2)\sinh2K_{1,2}e^{2K_{1,2}}]^{-1/2}, \;\;\; 
\tanh^2h^*_{1,2}=\tanh^2(h/2)\frac{\beta_{1,2}\exp(2K_{1,2})}{\cosh^2K^*
_{1,2}}, 
\end{eqnarray}
where $\alpha(h,x)=\tanh^2(h/2)(1+\cos x)/(\sin x)$ and 
$K_{1,2}={\beta}J_{1,2}, \;\; h={\beta}H, \;\; \beta=1/k_{B}T$ , 
where $T$ denotes temperature and $k_{B}$ - the Boltzman constant. 

\section*{Acknowledgements}
I am grateful to Dr. H. Makaruk for her help in preparation of the final 
form of this paper.

This paper was supported by the KBN grant $N^o$ {\bf 2 P03B 088 15}.

\newpage \pagestyle{empty}

\section{Captions for ilustrations}
Fig. 1 \,\,\, "Geometry" of transposition relations for $\alpha$-- and
$\beta$-- operators:

* --- $\alpha$-- operator

$\times$ --- $\beta$-- operator

\end{document}